\documentclass[prl,aps,showpacs,superscriptaddress,reprint]{revtex4-1}
\usepackage{amsmath,amssymb,graphicx}
\usepackage[final]{hyperref}
\usepackage{amsmath}
\usepackage{bbm}
\usepackage{bm}
\usepackage{braket}
\usepackage{float}
\usepackage[11pt]{moresize}
\usepackage{mathtools}
\usepackage[caption=false,labelformat=simple]{subfig}

\usepackage{placeins}

\hypersetup{
	colorlinks=true,       
	linkcolor=blue,          
	citecolor=blue,        
	filecolor=magenta,      
	urlcolor=blue   
}

\usepackage{setspace} 

\begin{document}
	
\title{An analytically simple and computationally efficient Gaussian beam mode-decomposition approach to classical diffraction theory}	

\author{Zhihao Xiao}
\email[xiaozhihao@hotmail.com]{}
\author{R. Nicholas Lanning}
\affiliation{Hearne Institute for Theoretical Physics, and Department of Physics $\&$ Astronomy, Louisiana State
	University, Baton Rouge, Louisiana 70803, USA}

\author{Mi Zhang}
\author{Irina Novikova}
\author{Eugeniy E. Mikhailov}
\affiliation{Department of Physics, College of William $\&$ Mary, Williamsburg, Virginia 23187, USA}

\author{Jonathan P. Dowling}
\affiliation{Hearne Institute for Theoretical Physics, and Department of Physics $\&$ Astronomy, Louisiana State
	University, Baton Rouge, Louisiana 70803, USA}
\affiliation{NYU-ECNU Institute of Physics at NYU Shanghai, Shanghai 200062, China}
\affiliation{CAS-Alibaba Quantum Computing Laboratory, USTC, Shanghai 201315, China}
\affiliation{National Institute of Information and Communications Technology, 4-2-1, Nukui-Kitamachi, Koganei, Tokyo 184-8795, Japan}

\begin{abstract}
We present a method of Gaussian-beam-mode decomposition to calculate classical diffraction of optical beams by apertures. This method offers a entirely different approach to examine the classic problem. Although our method is based on a very straightforward setup, it is surprisingly effective. We validate our method by comparing its results with those of Kirchhoff's full diffraction formula. Not only does our method have a simple and organized analytical framework, it also offers significant computational advantage.
\end{abstract}	

\maketitle 

\textit{Introduction.}---Traditional scalar diffraction theory for optical beams is based on the Huygens–-Fresnel principle and Kirchhoff's diffraction formula. There have been numerous studies \cite{jackson2007classical, born2013principles, bouwkamp1954diffraction, sommerfeld1954lectures} on the subject. In spite of some recent developments \cite{holmes1972parametric, mahajan1986uniform, nourrit2001propagation, drege2000analytical, dickson1970characteristics, holmes1972parametric, schell1971irradiance, campbell1969near}, the approach to classical diffraction is still based on these same principles. Scenarios under conditions such as a special field depth or a special symmetry in the aperture and source field have been studied \cite{campbell1987fresnel, lenz1996far, urey2004spot, tanaka1985field, lu1995focusing, burch1985fresnel, sheppard1992diffraction, mata2001diffraction, overfelt1991comparison, olaofe1970diffraction, belland1982changes}, but the applications are not as general as Kirchhoff's diffraction formula, and accuracy in the calculations needs improvement.

To simplify the calculation, the diffracted field is divided into the near field (known as Fresnel diffraction) or the far field (known as Fraunhofer diffraction), so that certain approximations can be applied. This arrangement is far from ideal, in the sense that we trade one problem with a complicated calculation for two problems, which require separate, simpler calculations. Additionally, there is no simple solution for the intermediate field between the near and the far field, and there is no clear picture describing how the near or the far field would transition into the other.

We propose a different approach for examining the diffraction problem, which is based on Gaussian-beam mode decomposition. 
While previous studies \cite{vciegis2002tool, trappe2003gaussian, brown2016fast} have been carried out on special cases where certain symmetry or other restrictions on the optical beams or apertures are assumed, here we investigate the general case. Our method, briefly discussed in Ref.~\cite{xiao2017iris}, takes advantage of the fact that spatial modes of a Gaussian beam, Laguerre-Gaussian (LG) modes and Hermite-Gaussian (HG) modes, each form a complete orthonormal basis in any given plane perpendicular to the beam axis. 
As a consequence, any source amplitude diffracted through an aperture can be expressed as a linear superposition of LG or HG modes. From now on we mainly focus on the LG modes, but for HG modes similar results can be derived.

\textit{Gaussian-beam-mode decomposition method.}---Gaussian-beam modes are the solutions that satisfy the free-space Maxwell's equations within the paraxial approximation. Specifically they are called Laguerre-Gaussian (LG) modes in cylindrical coordinates and Hermite-Gaussian (HG) modes in Cartesian coordinates. 
The exact mathematical expression of LG modes is \cite{Siegman_book}.
\begin{equation} \begin{split}
{u}_{ l,p}(r,\phi,z)=&\frac{C^{\rm{LG}}_{lp}}{w(z)}\left(\frac{r \sqrt{2}}{w(z)}\right)^{|l|}\exp\left(-\frac{r^2}{w^2(z)}\right)\\
& L_p^{|l|} \left(\frac{2r^2}{w^2(z)}\right) 	\exp\left( - i k \frac{r^2}{2 R(z)}\right)\exp( - i k z)\\
&\exp(i l \phi)\exp\left[i(2p+|l|+1)\zeta(z)\right],
\end{split}\end{equation}
where $r$, $\phi$ and $z$ are cylindrical coordinates; $l$ and $p$ are the azimuthal and radial indices, which are integers; $p \geqslant 0$; $C^{\rm{LG}}_{lp}=\sqrt{\frac{2}{\pi} \frac{p!}{(|l|+p)!}}$ is a normalization constant; 
$L_p^{|l|}$ is the associated Laguerre polynomial; $\lambda$ is the wavelength; $k=2 \pi/\lambda$ is the wave number;
$w(z)=w_0 \sqrt{1+(\frac{z}{z_R})^2}$ is the beam waist; $w_0$ is the beam waist at the beam focus;  $z_R=\frac{\pi w_0^2}{\lambda}$ is the Rayleigh range; $R(z)=z[1+(\frac{z_R}{z})^2]$ is the radius of curvature; $\zeta(z)=\arctan (\frac{z}{z_R})$ is the Gouy phase. Along the beam axis the beam waist will become wider or narrower, while the shapes of the intensity profiles remain similar. Note that we have chosen the $z$ axis to be the beam axis and $z=0$ to be the beam focus. 

\begin{figure*}
	\centering
	\subfloat[\label{fig:0214PlotKir22InputRectanAper0zR}]{
		\includegraphics[width=0.343\columnwidth]{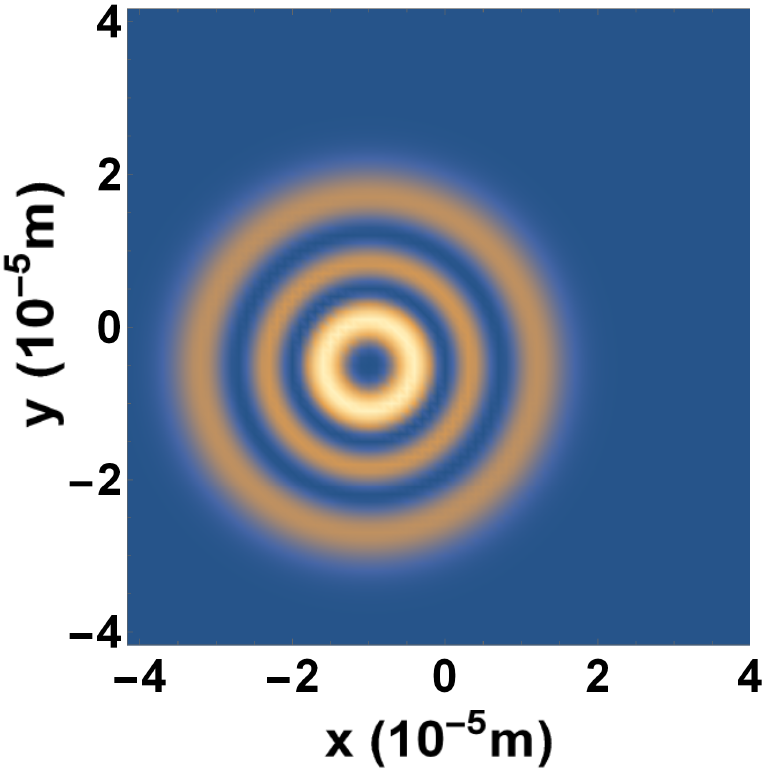}
	}
	\subfloat[\label{fig:0225UHG005PlotGauDecom22InputRectanAper0zR_Trim}]{
		\includegraphics[width=0.315\columnwidth]{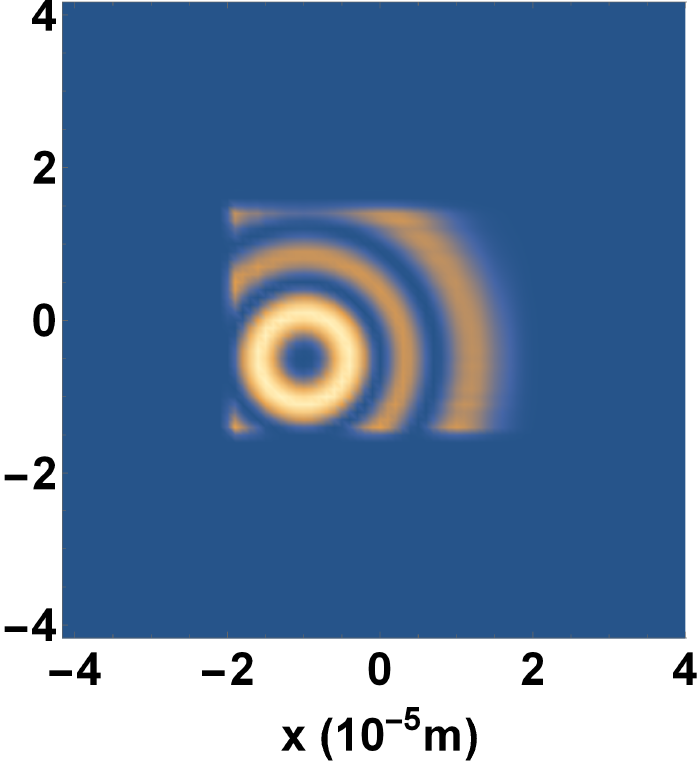}
	}
	\subfloat[\label{fig:0214PlotKir22InputRectanAper03zR_Trim}]{
		\includegraphics[width=0.312\columnwidth]{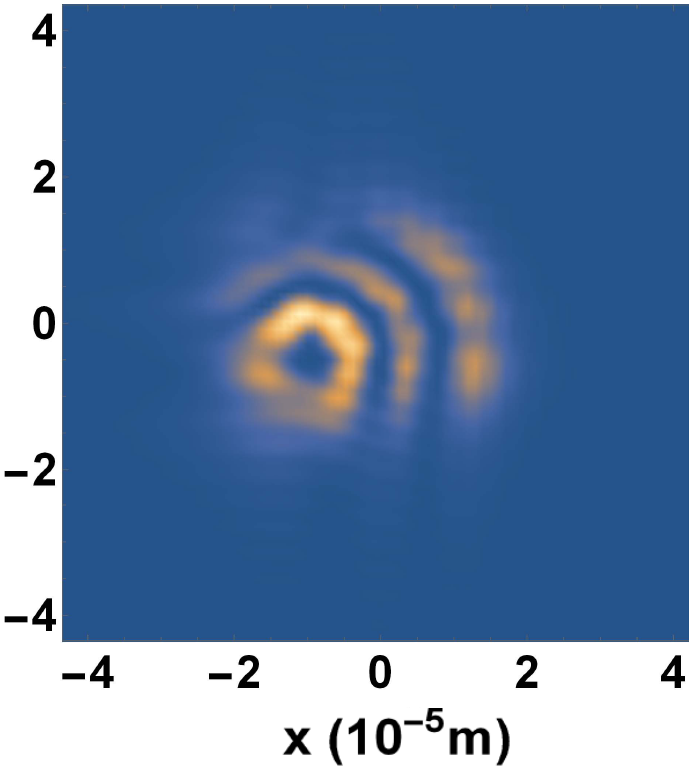}
	}
	\subfloat[\label{fig:0225UHG005PlotGauDecom22InputRectanAper03zR_Trim}]{
		\includegraphics[width=0.312\columnwidth]{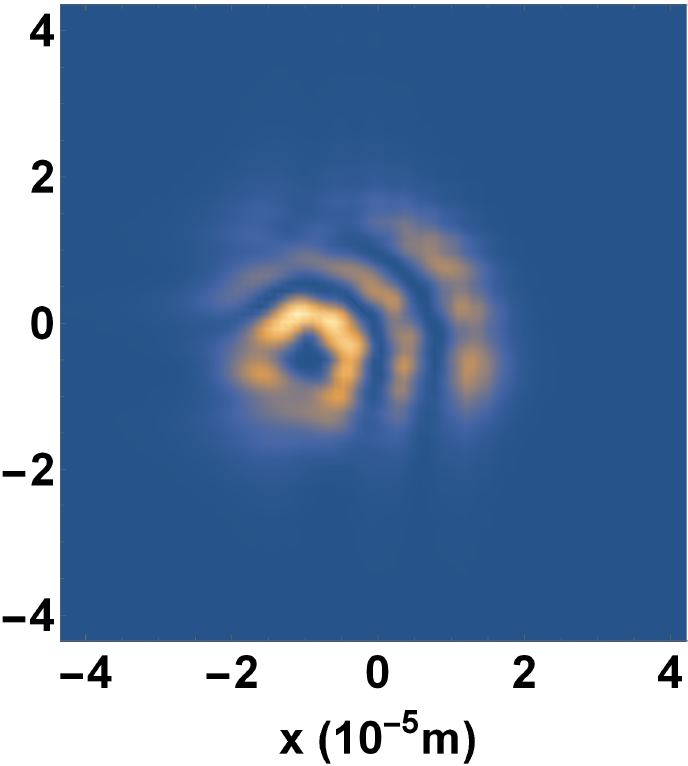}
	}
	\subfloat[\label{fig:0214PlotKir22InputRectanAper2zR_Trim}]{
		\includegraphics[width=0.318\columnwidth]{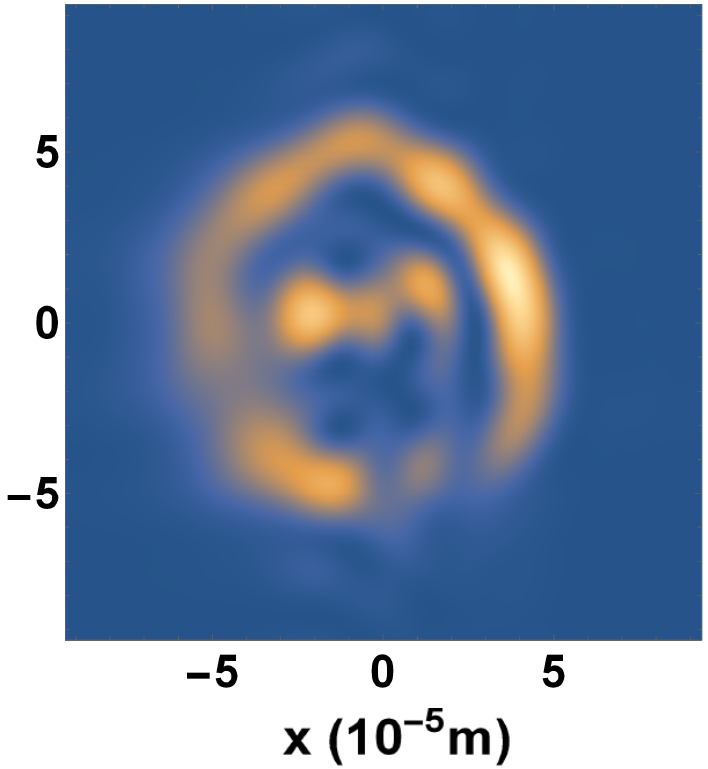}
	}
	\subfloat[\label{fig:0225UHG005PlotGauDecom22InputRectanAper2zR_Trim}]{
		\includegraphics[width=0.320\columnwidth]{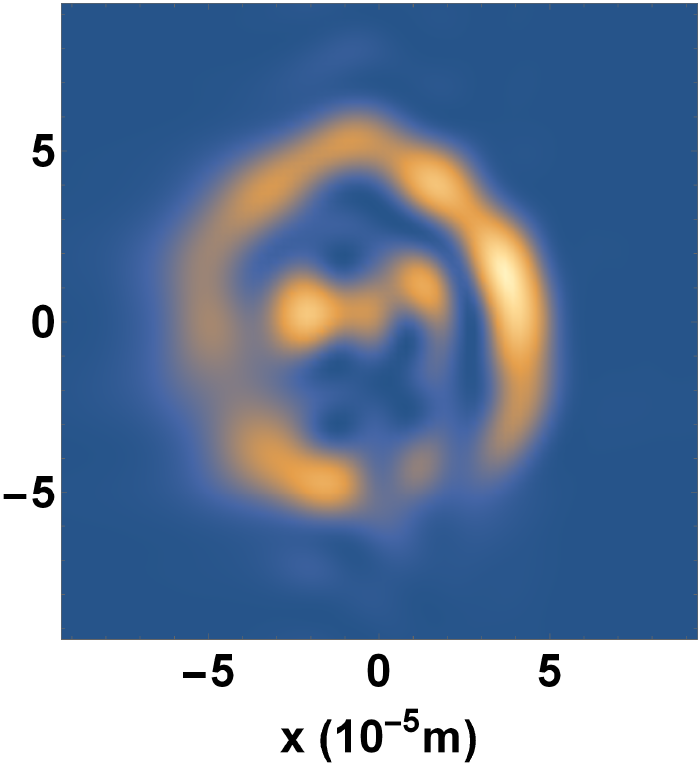}
	}
	\caption{\label{fig:0214PlotKirVsGauDecom22InputRectanAper}
		Intensity profiles in selected cross-sections.  An off-centered $l=2, p=2$ LG mode with wavelength $\lambda=796\text{nm}$ is used as the source field. It is traveling in positive $z$ direction and focused at ($x=-10^{-5}\text{m}, y=-0.5\times10^{-5}\text{m}, z=0\text{m}$) where the beam waist $w_0=10^{-5}\text{m}$ making its Rayleigh range $z_R=\frac{\pi w_0^2}{\lambda}=3.95\times10^{-4}\text{m}$. (a) shows its intensity profile right before it goes through the aperture which has an rectangular opening region $S=\{-2\times10^{-5}\text{m}<x<2\times10^{-5}\text{m}, -1.5\times10^{-5}\text{m}<y<1.5\times10^{-5}\text{m}, z=0\text{m}\}$. (b), (d), (f) show the intensity profiles in $z=0^+, 0.3z_R, 2z_R$ planes calculated using Gaussian-mode decomposition method. (c), (e) show the intensity profiles in $z=0.3z_R, 2z_R$ planes calculated using Kirchhoff's formula. Obviously (d) matches (c), and (f) matches (e). Also if we take the source field intensity pattern (a) and truncate it with aperture opening region $\mathbbm{S}_I$, the result matches (b). For the Gaussian-mode decomposition figures we have used $40\times40$ HG modes with a waist size of $5\times10^{-6}\text{m}$. For more information see \textit{Application to an Example with No Symmetry} on page 5.}
\end{figure*}

Also, the three parameters (the beam focus position, the wave number $k$ and the beam waist at the beam focus $w_0$) determine the entire LG mode orthonormal basis. Changing any one of the three parameters will give a different basis. We always have the freedom of choosing a particular set of the three parameters, in other words a particular LG mode basis that best suits our needs.

We can expand any optical beam propagating along the $z$ axis in terms of a superposition of LG modes, 
\begin{equation}\label{eqn:DiffractedAmpLG}
u(r,\phi,z)=\sum_{l,p}B_{l,p}\times{u}_{ l,p}(r,\phi,z).
\end{equation}

The orthonormality condition of the LG modes is,
\begin{equation}\label{eqn:orthogonalityLG}
\int\displaylimits_{\mathbbm{S}: z=z_0} u^*_{ l,p}(r_0,\phi_0,z_0) \times u_{ l',p'}(r_0,\phi_0,z_0) dS  = \delta_{ ll'} \delta_{ pp'} .
\end{equation}
It is important to notice that the above orthonormal condition only holds if the integration surface $\mathbbm{S}$ in Eq.~(\ref{eqn:orthogonalityLG}) is the entire $z=z_0$ plane, which can be any entire plane perpendicular to the beam axis. As a consequence, when a beam propagates through free space, different LG modes remain orthogonal. 

\textit{Analysis for a general aperture.}---When a classical beam passes through an aperture, the orthogonality between different LG modes breaks down and coefficients $B_{l,p}$ generally change. 

The orthonormality condition of the Gaussian modes demands the integration surface in Eq.~(\ref{eqn:orthogonalityLG}) to be an entire flat plane perpendicular to the beam axis. This is not particularly helpful if we want to examine the diffraction through apertures with a curved surface or not perpendicular to the beam axis. We consider the integration surface in Eq.~(\ref{eqn:orthogonalityLG}) to be a general, entire surface, where neither orthogonality nor normality holds.

The underlying principle of the classical diffraction is its quantum version \cite{xiao2017iris}, where $B_{l,p}$ are interpreted as coherent state amplitudes.

We now consider a general aperture, which is not necessarily perpendicular to the beam axis nor flat. We model the diffraction problem as follows. Let $\mathbbm{S}_A$ be the surface of the aperture where the optical beam is absorbed and $\mathbbm{S}_I$ be the surface of the opening of the aperture, which is illuminated by an amplitude distribution $u_I(r,\phi,z)$. $\mathbbm{S}_A\cup\mathbbm{S}_I$ forms a continuous, entire surface stretching to infinity in all transverse directions. 
Let us arrange the diffracted field amplitude coefficients $B_{l,p}$ into a column vector: 
\begin{equation}\label{}
B=\begin{bmatrix}
B_{l_1,p_1}       & B_{l_2,p_2}  &   \dots & B_{l_3,p_3}  & B_{l_4,p_4}  &  \dots 
\end{bmatrix}^T, 
\end{equation}
$l_i, p_i$ are row indexes.

Together with the Gaussian modes, the diffracted field amplitude is determined by $B$ which can be derived to be given by
\begin{equation}\label{eqn:CoefficientLG}
B=\mathbbm{M}^{-1} C,
\end{equation}
where $\mathbbm{M}$ is a matrix whose elements are given by
\begin{equation}\label{eqn:CoefficientLG_M}
\mathbbm{M}_{l,p;l',p'}=\int\displaylimits_{\mathbbm{S}_A\cup\mathbbm{S}_I} u^*_{l,p}(r,\phi,z) \times u_{l',p'}(r,\phi,z) dS_{\perp},
\end{equation}
where $l, p$ are row indexes, $l', p'$ are column indexes and $dS_{\perp}$ is the infinitesimal surface projected on the transverse plane. The vector $C$ is a column vector whose elements are given by
\begin{equation}\label{eqn:CoefficientLG_C}
C_{l,p}=\int\displaylimits_{\mathbbm{S}_I} u^*_{ l,p}(r,\phi,z) \times u_I(r,\phi,z) dS_{\perp}.
\end{equation}

Let us suppose an optical beam with a superposition of Gaussian modes propagates through free space. This amounts to a fictitious surface of $\mathbbm{S}_I$, which is illuminated by the superposition of Gaussian modes themselves. Since the free space has no absorption, $\mathbbm{S}_A=\emptyset$. With Eqs.~(\ref{eqn:CoefficientLG}--\ref{eqn:CoefficientLG_C}), we can verify that regardless how we construct the shape of $\mathbbm{S}_I$, so long it stretches to infinity in every perpendicular direction, the coefficients of the Gaussian modes remain the same, as they are supposed to. From this seemingly trivial scenario, we can see that all Eqs.~(\ref{eqn:CoefficientLG}--\ref{eqn:CoefficientLG_C}) are needed for a general aperture.

The most common situation is that the aperture is confined in a flat plane perpendicular to the beam axis. In this case, $\mathbbm{M}$ and its inverse $\mathbbm{M}^{-1}$ simplify to the identity matrix, and $B$ simplifies to $C$, and mode coefficients can be calculated from Eq.~(\ref{eqn:CoefficientLG_C}) alone. This gives us a new and intuitive way to interpret Eq.~(\ref{eqn:CoefficientLG}): $\mathbbm{M}^{-1}$ can be seen as the correction to $C$, due to the fact that orthonormal condition of Gaussian modes needs to be corrected when an aperture surface is neither flat nor perpendicular to the beam axis.

We have shown that our method offers a different way of calculating the diffracted field amplitude by making use of Eqs.~(\ref{eqn:DiffractedAmpLG}, \ref{eqn:CoefficientLG}) instead of Kirchhoff's diffraction formula.
Our method is developed based on the principle that the source beam and the diffracted beam should have the same boundary field amplitude at the aperture. This is different from Kirchhoff's diffraction formula, which is based on Huygens-–Fresnel principle.

Kirchhoff's diffraction formula takes the form of
\begin{equation} \begin{split}\label{eqn:DiffractedAmpKirchhoff}
u(r,\phi,z) = -\frac{i}{2\lambda} \int\displaylimits_{\mathbbm{S}_I} u_I(r',\phi',z') \frac{e^{ik\rho}}{\rho}(\cos\chi_1+\cos\chi_2)dS',
\end{split}\end{equation} 
where $\rho$ is the distance between the position of the diffracted field $(r,\phi,z)$ and the position of illumination field $(r',\phi',z')$, $\chi_1$ is the angle between aperture opening surface normal vector at $(r',\phi',z')$ and beam axis $z$, and $\chi_2$ is the angle between aperture opening surface normal and vector from $(r',\phi',z')$ to $(r,\phi,z)$.
(a) Our method has a neat and clear mathematical structure---Eqs.~(\ref{eqn:DiffractedAmpLG}, \ref{eqn:CoefficientLG}). It also has a simpler and more efficient computational formalism. The coefficients of the Gaussian modes $B_{l,p}$ are computed via Eq.~(\ref{eqn:CoefficientLG}) that are integrals. Once they are calculated, they can be used again and again to calculate the diffracted field at any position, using Eq.~(\ref{eqn:DiffractedAmpLG}). Calculating the amplitude in different positions in space simply means repeatedly doing the summations. This can be done very efficiently. 
(b) 
For Kirchhoff's diffraction formula, near- and far-field approximations generally result in different outcomes, causing difficulties if both near and far field need to be examined. 
On the other hand, our method offers one unified and already simple computing framework, which applies to all near, far, and intermediate fields.
(c) 
For our method, to yield completely accurate results, all (infinite) orders of Gaussian modes in Eqs.~(\ref{eqn:DiffractedAmpLG}, \ref{eqn:CoefficientLG}) must be accounted for, which means $l \in (- \infty, + \infty) $ and $p \in (0, + \infty)$. But the coefficients for higher-order modes usually drop off rapidly. According to Eq.~(\ref{eqn:CoefficientLG}), every mode coefficient is calculated individually, meaning the calculation of each coefficient is independent of the others. If higher accuracy is needed, we can simply calculate additional higher-order coefficients on demand and add the amplitude to the existing lower-order mode amplitudes. In other words, we can make use of the lower-accuracy result to obtain the higher accuracy result.

\textit{Demonstration of the validity.}---We now compare our method to Kirchhoff's diffraction formula. Let us examine a simple case in which a circular aperture with radius $a$ is placed at the $z=z_I$ plane and centered on the beam axis, and the aperture is illuminated by a plane wave traveling along the beam axis with normalized amplitude and wavelength $\lambda$. 
The surface in Eq.~(\ref{eqn:CoefficientLG}) now has the form of $\mathbbm{S}_I=\{r<a; z=z_I\}$.

Kirchhoff's diffraction formula (in cylindrical coordinates) takes the form of \cite{Siegman_book}
\begin{equation} \begin{split}\label{eqn:DiffractedAmpKirchhoffCylindrical}
u(r,z)=& i2\pi N e^{-i\pi N(r/a)^2}\\
&\int\limits_{0}^{a} \frac{r_0 u_I e^{-i\pi N(r/a)^2}}{a^2} J_0(\frac{2\pi N r r_0}{a^2})dr_0,
\end{split}\end{equation}	
where the source field (plane wave) amplitude $u_I(r,\phi,z_I)$ is set to be uniformly $u_I$. The diffracted field amplitude at plane $z$ and radial distance $r$ is $u(r,z)$. $N\equiv\frac{a^2}{(z-z_I)\lambda}$ is the Fresnel number and $J_0$ is the zeroth order Bessel function. 
The region where the Fresnel number $N<<1$ is considered as far-field diffraction, and Kirchhoff's diffraction formula simplifies to,
\begin{equation} \begin{split}
u(r,z)=i\pi N e^{-i\pi N (r/a)}\frac{2 J_1(2\pi N r/a)}{2\pi N r/a},
\end{split}\end{equation}	
where $J_1$ is the first order Bessel function. This result is the well-known Airy pattern. The region where the Fresnel number $N>>1$ is considered near-field diffraction.
We can see that there are a very limited number of situations where we can simplify Kirchhoff's diffraction formula, as there is no simplification even for this uniformed source field and highly symmetrical aperture in the intermediate field.

For all non-zero $l$, $B_{l,p}=0$, since the source-field amplitude and the aperture are both cylindrically symmetric. Therefore only $l=0$ modes contribute to diffracted field amplitude. Furthermore, in the numerical simulation we need to set a maximum $p$ index $p_{\text{max}}$ and all modes exceeding $p_{\text{max}}$ will be ignored. Therefore the result is reduced to,
\begin{equation} \begin{split}
u_{\text{eff}}(r,\phi,z)=\sum_{p=0}^{p_{\text{max}}}B_{0,p}\times{u}_{ 0,p}(r,\phi,z).
\end{split}\end{equation}

\begin{figure*}
	\centering
	\subfloat[\label{fig:0728PlainWave_10W0_P80&Kir_10N}]{
		\includegraphics[width=0.380\columnwidth]{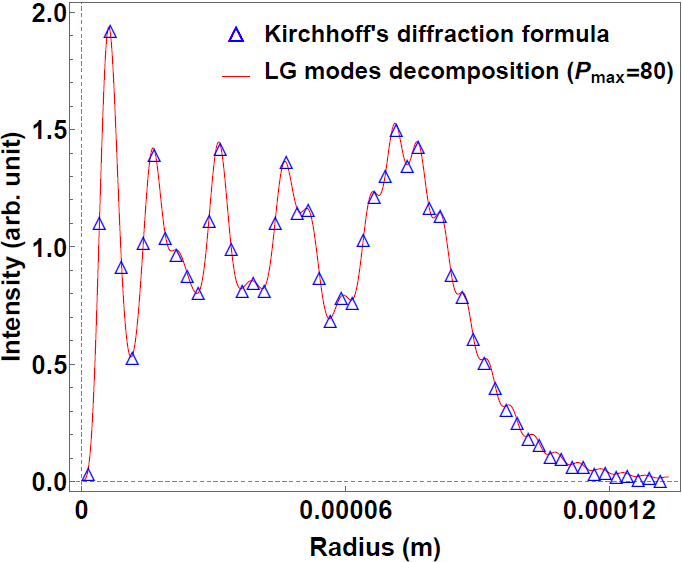}
	}
	\subfloat[\label{fig:0728PlainWave_10W0_P80&Kir&Airy_1N_Trim}]{
		\includegraphics[width=0.360\columnwidth]{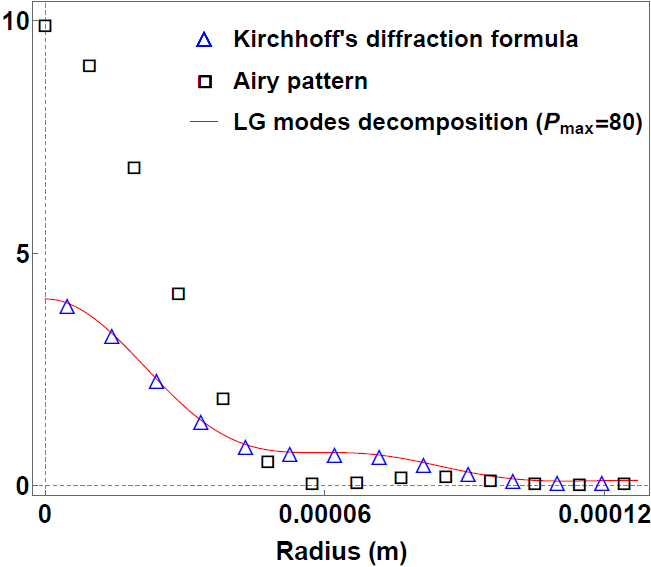}
	}
	\subfloat[\label{fig:0728PlainWave_10W0_SmallR_P80&Kir&Airy_01N_Trim}]{
		\includegraphics[width=0.370\columnwidth]{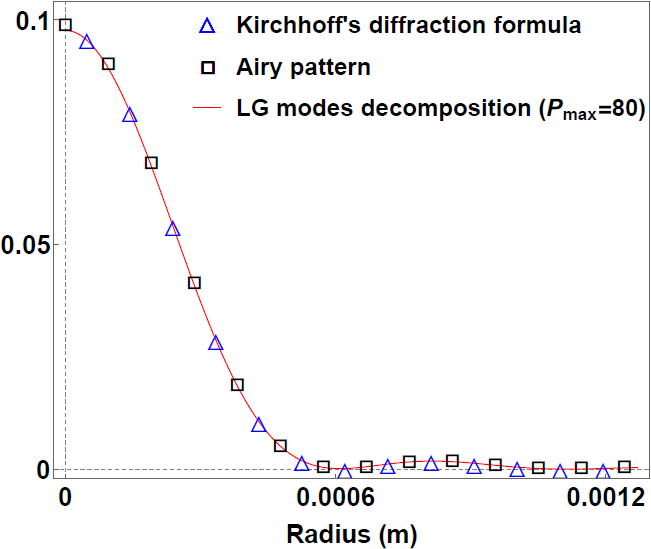}
	}
	\subfloat[\label{fig:0728PlainWave_10W0_LargeR_P80&Kir&Airy_01N_Trim}]{
		\includegraphics[width=0.410\columnwidth]{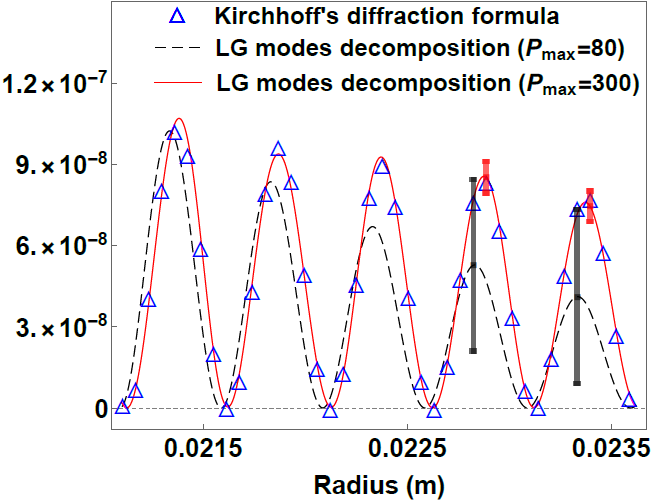}
	}
	\subfloat[\label{fig:ConvergingPowerRatio}]{
		\includegraphics[width=0.400\columnwidth]{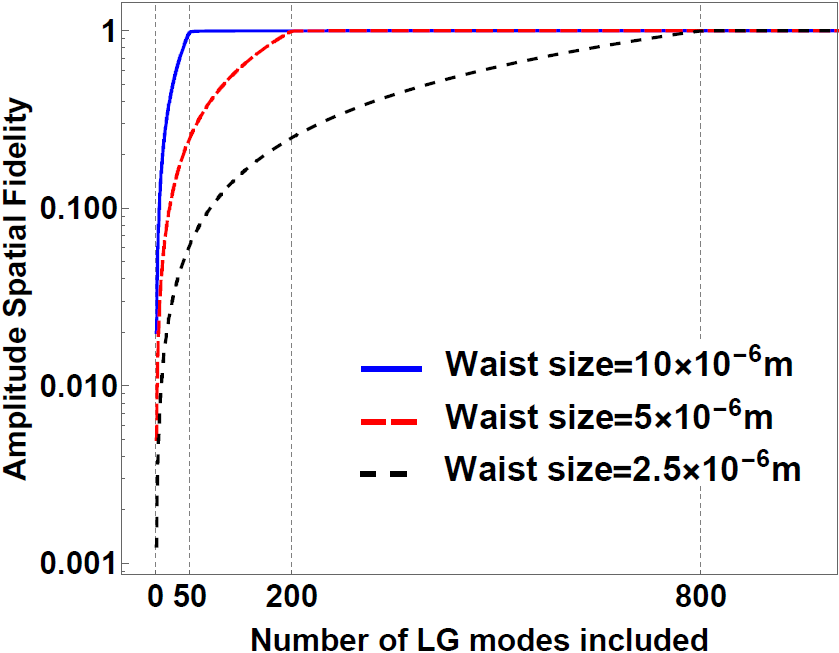}
	}
	
	\caption{\label{fig:0728PlainWave_10W0}
		(a)--(d): intensity vs. radius for plain wave diffraction through a circular aperture in (a) the near field where $N=10$ and (b) the intermediate field where $N=1$. (c) shows intensity in the far-field, near-axis region while (d) shows the far-field, far-axis region, both of which have $N=0.1$. (d) also shows error bars for several radius positions and different $p_{\text{max}}$ curves. The wavelength is $\lambda = 796\text{nm}$ and aperture radius is $a=10^{-4}\text{m}$. For the LG modes used in this simulation, we choose the focus to be at the center of the aperture where the beam waist is $w_0=10^{-5}\text{m}$. (e) shows $F(u_{\text{eff}}, u)$ vs. $p_{\text{max}}$. Different curves represent LG modes basis with different chosen waist sizes. With a smaller waist size, a larger number modes must be included to cover the regime where $F(u_{\text{eff}}, u)$ converges to 1 sharply. Notice this number, which is roughly where the sharp converging regime ends, is quadrupled when the waist size is halved.
		}
\end{figure*}

Since the diffracted field has cylindrical symmetry, we need to only examine the intensity in terms of radius in near, intermediate and far field regions, as shown in Fig.~\ref{fig:0728PlainWave_10W0_P80&Kir_10N}--\ref{fig:0728PlainWave_10W0_LargeR_P80&Kir&Airy_01N_Trim}. In all these regions we can see that our method fits very well with Kirchhoff's diffraction formula. Notice in the far-field, far-axis region (shown in Fig.~\ref{fig:0728PlainWave_10W0_LargeR_P80&Kir&Airy_01N_Trim}) the result from our method with cutoff $p_{\text{max}}=80$ is deviating from the Kirchhoff's formula result (Airy rings). The further we go to the far-axis region (larger radius) the more inaccurate the $p_{\text{max}}=80$ result becomes. This inaccuracy can be easily improved by including more higher-order LG modes, as the $p_{\text{max}}=300$ result shows. A similar phenomenon is observed in the far-axis region of near field and intermediate field as well: the further from the beam axis, the more modes are needed to be accurate. The explanation of this phenomenon is two fold: (a) lower order LG or HG modes concentrate a large part of their intensity near the beam axis, while the higher-order modes have a much more spread-out intensity distribution in both near-axis and far-axis. Therefore, lower-order modes are the dominating factor in the near-axis region, while higher-order modes must be included in the far-axis. (b) In general, the far-axis intensity is lower than near-axis, as one can see from the ranges of vertical axis of  Fig.~\ref{fig:0728PlainWave_10W0_SmallR_P80&Kir&Airy_01N_Trim} and Fig.~\ref{fig:0728PlainWave_10W0_LargeR_P80&Kir&Airy_01N_Trim}. Therefore inaccuracy is more apparent in far-axis than in near-axis, making the far-axis result more sensitive to error. 

Though the Kirchhoff integral simplification to the Airy pattern is accurate in the far field (Fig.~\ref{fig:0728PlainWave_10W0_SmallR_P80&Kir&Airy_01N_Trim}), it performs poorly in the intermediate field (Fig.~\ref{fig:0728PlainWave_10W0_P80&Kir&Airy_1N_Trim}). It performs even worse in the near field (which is not shown in Fig.~\ref{fig:0728PlainWave_10W0_P80&Kir_10N} because it is too inaccurate). This provides evidence for our point that 
our method has the same range of application as the full Kirchhoff's formula.

The essence of our method is to represent the diffracted field in a LG (or HG) basis. If all LG modes are included, the method gives a completely accurate result. However, for practical reasons, only limited LG modes ($l=0, p\leq p_{\text{max}}$) are included, and $u_{\text{eff}}(r,\phi,z)$ is employed to approximate $u(r,\phi,z)$. To examine the closeness between the amplitude distributions of $u_{\text{eff}}(r,\phi,z)$ and $u(r,\phi,z)$ across any entire perpendicular diffraction plane $z=z_1$, we introduce an amplitude spatial fidelity defined as 
\begin{equation}\begin{split}\label{eqn:amplitude_spatial_fidelity}
F(u_{\text{eff}}, u)\equiv\frac{|\int\displaylimits_{\mathbbm{S}: z=z_1}u^*_{\text{eff}}(r,\phi,z) u(r,\phi,z) dS|^2}{\int\displaylimits_{\mathbbm{S}: z=z_1}u^*_{\text{eff}} u^{}_{\text{eff}} dS \times \int\displaylimits_{\mathbbm{S}: z=z_1}u^* u dS},
\end{split}\end{equation}
which is closer to one if $u_{\text{eff}}(r,\phi,z)$ resembles $u(r,\phi,z)$ more. This fidelity does not depend on the position of $z_1$, therefore we can examine the plane immediately after the aperture $(z_1=z_I^+)$. With our method, the calculation of amplitude spatial fidelity is conveniently simplified to: 
$F(u_{\text{eff}}, u)=\sum\displaylimits_{p=0}^{p_{\text{max}}}|B_{0,p}|^2/\int\displaylimits_{0<r<a} |u_I|^2 2\pi rdr$. The fidelity cannot be so conveniently calculated with Kirchoff's diffraction formula.

We have the freedom of choosing the Gaussian mode basis.
However, some Gaussian mode bases work better than others. There is no black-or-white rule regarding the choice of basis, 
so we only give some general guidelines here. To be able to better depict the variation of the diffracted field across various transverse positions, we should choose a basis with a small waist. In addition, in order to better depict the variation of the diffracted field along the beam axis, especially in the near field, we should choose a basis with a smaller Rayleigh range, which also requires a smaller waist. However, the smaller the waist we choose, the more modes we will generally need to include. To illustrate this, we can examine $F(u_{\text{eff}}, u)$ versus the number of modes included. As the number of modes increases in one particular basis, $F(u_{\text{eff}}, u)$ converges to one, typically in two distinct regimes: sharp converging and long tail regimes, shown in Fig.~\ref{fig:ConvergingPowerRatio}. Once a basis is chosen, sufficient modes need to be included to at least cover the sharp converging regime, otherwise a significant part of the amplitude will be unaccounted for. For this to happen, we have found the number of modes needed is inversely proportional to the square of waist size. The area one Gaussian mode can cover is proportional to the square of waist size. If we use a smaller waist size, to fit in the same illuminated area, the number of modes needed must increase.

On one hand, $F(u_{\text{eff}}, u)$ is useful for assessment of the accuracy of $u_{\text{eff}}(r,\phi,z)$ globally across any perpendicular diffraction plane. On the other hand, to assess the accuracy of our effective diffraction field intensity calculation locally in any specific position, we can use the following error estimation,
\begin{equation}\begin{split}\label{eqn:ErrorIntensity}
\text{Err}(r,\phi,z;p_{\text{max}})=&\left|\int_{0<r<a} |u_I|^2 2\pi rdr-\sum\displaylimits_{p=0}^{p_{\text{max}}}|B_{0,p}|^2\right|\\
&\times\max\{|{u}_{ 0,p}(r,\phi,z)|^2\},\  p>p_{\text{max}},
\end{split}\end{equation}
which works well in the limit of $F(u_{\text{eff}}, u)\rightarrow1$. If, in a certain region, the error becomes comparable to the intensity, it suggests that the modes currently included are no longer adequate. The error bars in Fig.~\ref{fig:0728PlainWave_10W0_LargeR_P80&Kir&Airy_01N_Trim}, calculated with Eq.~(\ref{eqn:ErrorIntensity}), show that the error bars provide effective warning that $80$ modes are no longer sufficient in this far axis region, and by increasing $p_{\text{max}}$, from 80 to 300 in this case, we can reduce the error and achieve greater accuracy.


\textit{Application to an Example with No Symmetry.}---As mentioned previously, our method can be applied to any source field and any aperture. In the absence of symmetry, Kirchhoff's formula generally cannot be simplified and the power of our method becomes even more apparent. In this example, we are using a off-centered $l=2, p=2$ LG mode as the source field (shown in Fig.~\ref{fig:0214PlotKir22InputRectanAper0zR}) and a rectangular aperture centered on the beam axis. In this setup, there is no symmetry that is helpful for simplification of Kirchhoff's formula. We again compare the intensity patterns generated by our method and by Kirchhoff's formula, shown in Figs.~\ref{fig:0225UHG005PlotGauDecom22InputRectanAper0zR_Trim}--\ref{fig:0225UHG005PlotGauDecom22InputRectanAper2zR_Trim}.
Our method is validated again, as its results matches the results from Kirchhoff's formula. In each of the 2D plots, there are $100\times100$ data points. Due to the computational advantages of our method, generating  its plots takes only a small fraction of time as the plots with Kirchhoff's formula. Should we need to generate more data points, the superiority of our method becomes more apparent.

\textit{Conclusion.}---We have developed the Gaussian-mode decomposition method to calculate classical diffraction of optical beams through apertures of arbitrary shape. We have explained the setup of this method and demonstrate the effectiveness and efficiency of our method. We validate our method by comparing its result with that of Kirchhoff's diffraction formula. In addition, we have established a proper way of accessing accuracy. We have shown that not only our method has a simple and organized analytical framework, it also offers significant computational advantage over the traditional Kirchhoff's diffraction formula.

\textit{}The authors would like to acknowledge support from the the Air Force Office for Scientific Research, the Army Research Office, the Defense Advanced Projects Agency, the National Science Foundation, and the Northrop Grumman Corporation.


\bibliography{bibliography/bibliographyshort}

\end{document}